\documentclass[12pt]{iopart}
\usepackage{iopams}
\usepackage{graphicx}
\usepackage{psfrag}
\usepackage{dcolumn}
\usepackage{latexsym}
\usepackage{amssymb,latexsym,mathrsfs,color}
\usepackage{amsfonts}

\newcommand{\beq}{\begin{equation}}
\newcommand{\eeq}{\end{equation}}
\newcommand{\beqn}{\begin{eqnarray}}
\newcommand{\eeqn}{\end{eqnarray}}
\newcommand{\bearr}{\begin{array}}
\newcommand{\enarr}{\end{array}}

\newcommand{\ra}{\rangle}

\def\bea{\begin{eqnarray}}
\def\eea{\end{eqnarray}}
\def\ba{\begin{array}}
\def\ea{\end{array}}

\def\theta{\alpha}

\begin{document}
 
\title{Comments on ``Critical Study on the Absorbing Phase Transition in a
Four-State Predator-Prey Model in One Dimension''}
\author{P. K. Mohanty, Rakesh Chatterjee and Abhik Basu}
\address{Theoretical Condensed Matter Physics Division,\\ Saha Institute of
Nuclear Physics, Kolkata 700064, India.}

\def \rb {\rho^{s}_{_B}}
\def \ra {\rho^{s}_{_A}}
\def \rbt {\rho_{_B}}
\def \rat {\rho_{_A}}
\begin{abstract}
%

In  a recent article [arXiv:1108.5127]  Park has  shown
that  the four-state predator-prey  model  studied  earlier in  {\it J. Stat. Mech, L05001 (2011)}
belongs to  Directed Percolation (DP) universality class.  It was  claimed that 
predator density is not a reasonable order parameter, as there  are many absorbing states; a
suitably chosen order parameter  shows DP critical behaviour.  In this  article, we  argue that  
the configuration that does not have  any predator is the only dynamically accessible absorbing  
configuration, and the predator density too settles to DP critical exponents after a long transient.  
\end{abstract}


\maketitle

Systems having absorbing configurations may undergo a non-equilibrium phase 
transition \cite{AAPTBook}  from an  active to an  absorbing state.  
The critical behavior of these absorbing state phase transitions (APTs) \cite{DPBook}
depends on the the symmetry  of the order parameter and  presence of additional conservation laws. 
It has been conjectured \cite{DPconj} that in absence of any
special symmetry the APT belongs to the directed percolation (DP) 
universality class as long as the system has a single absorbing state.

Since  the coarse grained microscopic theory of DP, which is a  birth-death-diffusion 
process,  is based on a single component Reggeon field theory\cite{Reggeon}, critical behavior 
in presence of  additional field is expected to alter  the critical behavior.  The 
additional field may bring in multiple absorbing states  and/or additional  conservation 
laws. Presence of multiple absorbing states may \cite{inf_no} or may not 
 \cite{inf_yes} affect the universality. Coupling of order parameter to a conserved 
field too  lead to DP \cite{dd} or non-DP  \cite{CTTP} critical behavior.  
The models of directed percolation  with more than one species  
\cite{twospc}, which brings  in  additional  coarse grained fields,  
has also been studied \cite{Dickman:91}. The predator-prey cellular 
automaton models \cite{lipowska} in higher dimension 
too shows an APT to an absorbing (extinct) state which belongs to DP-class.
The role of additional fields in these models are not quite  well understood.

Recently we studied a  predator-prey model\cite{4SPP} on a  $(1+1)$-dimensional lattice,  where
each lattice site is either vacant, occupied by  a predator $A$, a prey $B$ or both (one $A$ and one $B$). 
In these  four state predator-prey  (4SPP) model  growth of preys  and   death  of predators 
occurs independently, whereas  death of a prey is  always   followed by instant birth of a predator.    
Based on the numerical  simulations and estimated critical exponents, we have suggested the
possibility of a new universality class. In particular, the decay of clusters  at 
the critical point  was found to be distinctly different from  those of DP.  However, in a 
recent article  Park \cite{Park} has suggested a  different scenario. It was claimed that  
the predator density $\rho_B$ can not be taken as a order parameter  as there are  infinitely many 
absorbing states. The transition  is found to be  in DP class,   when order parameter is chosen 
suitably. In this article, we show that  although there are many absorbing states, only one of them  
$( \rho_A=1, \rho_B=0)$ is dynamically accessible. In fact  the order parameter $\rho_B$  which was 
showing an apparently new  critical behavior, slowly  crosses over to DP.


For completeness, first let us define the model. On a one dimensional periodic lattice, 
each site is either vacant, or occupied by a
single particle $A$ (prey), or occupied by a single particle $B$ (predator) or
by both particles (co-existing $A$ and $B$);  correspondingly at each site 
$i$ we have  $s_i=0,1,2,3.$  Alternatively one may  describe the 
4-state predator-prey (4SPP) model considering  two separate branches, one for
$A$ and the other for $B$ particles, where particles living in one branch can 
not move to the other. 
We consider only the asymmetric case of the model which evolves according to the following 
dynamics, where $X$ denotes both presence and  absence of particles  in respective branches.

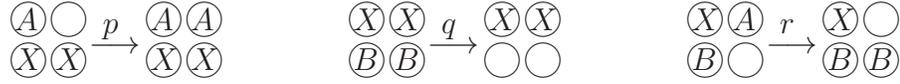
\begin{center}
\setlength{\unitlength}{.9mm}
\begin{picture}(100, 35)
\put(-10,30){\circle{5}}\put(-11,28.5){$ $}\put(-16,30){\circle{5}}\put(-18,28.5){$A$}
\put(-10,24){\circle{5}}\put(-12,22.5){$X$}\put(-16,24){\circle{5}}\put(-18,22.5){$X$}
\put(-7,25){$\longrightarrow$}\put(-5,28){$p$}
\put(10,30){\circle{5}}\put(8,28.5){$A$}\put(4,30){\circle{5}}\put(2,28.5){$A$}
\put(10,24){\circle{5}}\put(8,22.5){$X$}\put(4,24){\circle{5}}\put(2,22.5){$X$}

\put(40,30){\circle{5}}\put(38,28.5){$X$}\put(34,30){\circle{5}}\put(32,28.5){$X$}
\put(40,24){\circle{5}}\put(38,22.5){$B$}\put(34,24){\circle{5}}\put(32,22.5){$B$}
\put(43,25){$\longrightarrow$}\put(45,28){$q$}
\put(60,30){\circle{5}}\put(58,28.5){$X$}\put(54,30){\circle{5}}\put(52,28.5){$X$}
\put(60,24){\circle{5}}\put(59,22.5){$ $}\put(54,24){\circle{5}}\put(53,22.5){$ $}

\put(90,30){\circle{5}}\put(88,28.5){$A$}\put(84,30){\circle{5}}\put(82,28.5){$X$}
\put(90,24){\circle{5}}\put(89,22.5){$ $}\put(84,24){\circle{5}}\put(82,22.5){$B$}
\put(93,25){$\longrightarrow$}\put(95,28){$r$}
\put(110,30){\circle{5}}\put(109,28.5){$ $}\put(104,30){\circle{5}}\put(102,28.5){$X$}
\put(110,24){\circle{5}}\put(108,22.5){$B$}\put(104,24){\circle{5}}\put(102,22.5){$B$}
\end{picture}
\end{center}
\vspace*{-1.5 cm}

Although, it was not explicitly mentioned in our earlier report \cite{4SPP}, numerical simulations  of 
the model was carried out in  using exactly the  same state variables 
$b_i= 4s_i  + s_{i+1}$ mentioned in \cite{Park}. Clearly 
these bond variables $b_i= 0, 1\dots 15$ follow a dynamical rules,
\begin{eqnarray}
4 \mathop{\longrightarrow}^{p}5 ; 6\mathop{\longrightarrow}^{p} 7 ;12\mathop{\longrightarrow}^{p} 13; 
14\mathop{\longrightarrow}^{p} 15;\quad 
9\mathop{\longrightarrow}^{r}10; 13\mathop{\longrightarrow}^{r}14 \label{eq:1} \\
10 \mathop{\longrightarrow}^{q}0;11 \mathop{\longrightarrow}^{q}1;14\mathop{\longrightarrow}^{q} 4;15 \mathop{\longrightarrow}^{q}5;\label{eq:2}  
\end{eqnarray}
The neighbors are updated  along with $b_i$ as follows. For  dynamics (\ref{eq:1}) $b_{i+1}$  is increased 
by $4$ whereas  for   dynamics (\ref{eq:2})  $b_{i+1}$ and    $b_{i-1}$ are decreased by $8$ and $2$ respectively.
Clearly a bond $i$ is active when $b_i$ is either $4$ or $6$, or it is  greater than $8.$  It  was argued by 
Park \cite{Park}  that the  density of active  bonds  $\rho$,  the correct order parameter (as there are  
infinitely many absorbing states)  of the system, vanishes at the critical point $p_c=0.15381$ when 
$q=0.02$ and $r=0.9.$  For the same parameter values, we had  estimated earlier that predator density 
$\rho_B$  vanishes at $p_c^B=0.1484$.  This raises a question that   possibly  in the region $p_c^B<p\le p_c$, 
the system falls into an absorbing state which has isolated $B$s.  However   a configuration  with 
one or more  isolated $B$ can be absorbing only when there are   no $A$s.  Thus, at $p_c$ 
we must have $\rho_A=0.$  It was evident from Fig. 2(a) of  Ref. \cite{4SPP} that  
$\rho_A >0 ~~\forall p.$ This indicates that  $\rho_B$  must vanish at the same   value of $p,$ 
possibly at $p_c= 0.15381,$ as  estimated by Park \cite{Park}.

\begin{figure}[h]
\vspace*{1cm}
 \centering
\includegraphics[width=14cm]{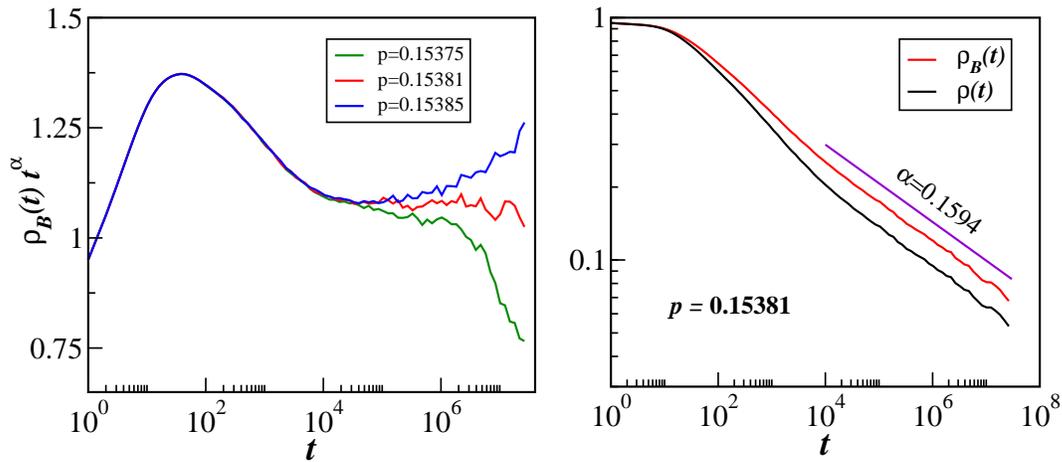}
\caption{(a) Plots of $\rho_B(t) t^\alpha$ vs $t$  for $p = 0.153 75, 0.153 81, 0.153 85$ with  
$q = 0.02$, $r= 0.9$ and $L=2^{20}$. Here we use DP critical exponent  $\alpha= 0.1594.$  (b) Both  $\rho(t)$ and 
$\rho_B(t)$ asymptotically decay  as $t^{- 0.1594}$  at  $p_c=  0.153 81$.}
\label{fig:pc}
\end{figure}

To check this we redo the Monte-Carlo simulation for larger system size $L=2^{20}$ 
and measured  $\rho_B(t)$   up to $t= 3\times 10^7$ MCS for  different values of $p,$ while 
keeping $q=0.02$ and $r=0.9.$ From Fig. \ref{fig:pc}(a)   it is evident that  $\rho_B(t)  t^\alpha$ 
shows a saturation  at $p_c= 0.153 81$  for $\alpha= \alpha_{DP}=0.1594.$  Figure 
\ref{fig:pc}(b) shows   log-scale plot of $\rho(t)$ and $\rho_B(t)$ at  $p_c;$  it is evident 
that  after a long transient   $\rho_B(t)$ decays  as $t^{-\alpha}$ with  $\alpha = 0.1594$, 
similar to $\rho(t).$   Such a  change in $\alpha$  to a lower  value was appearing as a saturation 
in $\rho_B$, leading to a lower estimate of the critical point.

\begin{figure}[h]
\vspace*{1cm}
 \centering
\includegraphics[width=13cm]{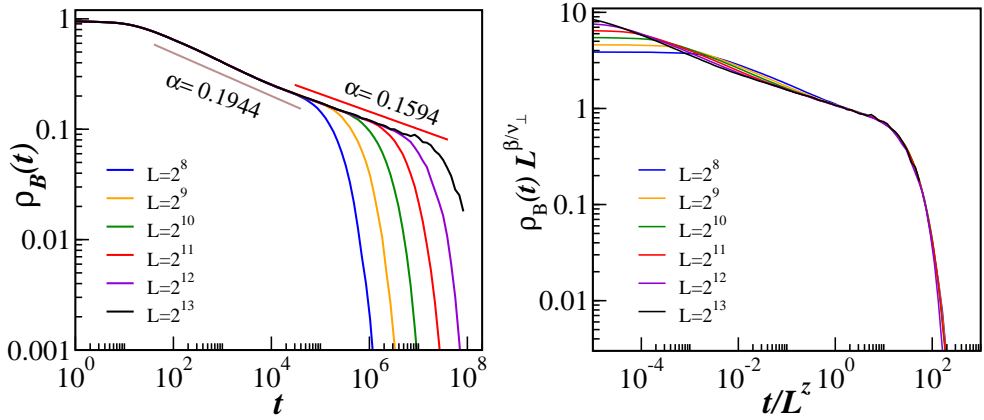}
\caption{(a)  Log-scale plots of $\rho_B(t)$ vs $t$  at criticality ($p=0.15381,$  
$q = 0.02$ and $r= 0.9$)for $L= 2^8, 2^9, 2^{10}, 2^{11}, 2^{12}$ and $2^{13}$.  The initial slope $\alpha=0.194$ crosses over to $\alpha_{DP}=0.1594$ as system size is increased. 
(b) The same data are collapsed according to Eq. (\ref{eq:rhot_L}) by using  DP critical 
exponents $\beta/\nu_\perp=0.252,$  and $z=1.580,$} 
\label{fig:z_nu}
\end{figure}

With this correct estimation of critical point, we  proceed to calculate  other critical exponents  
taking  $\rho_B$ as the order parameter. As for  finite system  of size $L$,
starting from a high  density of predators,  
$\rbt(t,L)$ decays as $t^{-\alpha}$, indicating a scaling form
\begin{equation}
\rbt(t, L) = L^{-\beta/\nu_\perp} { \tilde {\cal G}}(t/L^z),
\label{eq:rhot_L}
\end{equation}
at the critical point, where $z$ is the dynamical critical exponent. Thus, $\rbt(t)L^{\beta /\nu_\perp}$
 for different values of $L$  are expected to collapse to a single function when plotted
against $t/L^z$. Figure \ref{fig:z_nu}(a)  shows decay  of $\rbt(t)$ for different $L$  starting 
from  the configuration $\rho_A=1=\rho_B$;  clearly  small systems show an effective exponent 
$\alpha=0.194$  which  crosses 
over to $\alpha_{DP}$ as the system size is increased. True finite size effect of the 
critical point  sets in at a reasonably large $L.$  The data collapse according to 
Eq. (\ref{eq:rhot_L})  is observed in Fig. \ref{fig:z_nu}(b) where  we use  the DP exponents 
$\frac{\beta} {\nu_\perp} = 0.252$ and $z=1.580.$

In summary, although there are  many  absorbing configurations in  4SPP model, 
the numerical simulations suggests that only one of them  $(\rho_A=1,\rho_B=0)$ 
is dynamically accessible. The critical behavior of the absorbing transition 
can be  well described by taking $\rho_B$ as an order parameter. The earlier 
estimated values of the critical exponents vary slowly  with system size and 
settle to DP values. These studies truly emphasize the complications, 
 difficulties  and danger associated in  numerical determination of 
universality class, when the possibility of a long transient is not ruled out. 

{\it Acknowledgement :} PKM would like to acknowledge M. Basu and U. Basu for 
 doing independent  Monte-Carlo simulations  to confirm  the results.

\end{document}